\documentclass[onecolumn,
 preprint,
 nofootinbib,
 amsmath,amssymb,
 aps,
]{revtex4-2}

\usepackage{graphicx}
\usepackage{dcolumn}
\usepackage{bm}
\usepackage[colorlinks,citecolor=blue]{hyperref}
\usepackage{xcolor}

\usepackage{verbatim}
\usepackage{color}

\begin{document}
\preprint{PITT-PACC-2323}
\title{Axion Electrodyanmics in the Presence of Current Sources}
\author{Joshua Berger}
\author{Amit Bhoonah}
\date{\today}
\begin{abstract}
Axions are among the most sought-after candidates for dark matter.  In the ultralight regime, they could help alleviate puzzles in small scale cosmology.  Searches for a halo of axion dark matter rely on the electromagnetic response to a magnetic field.  In this work, we resolve a number of issues in the determination of this response by carefully solving Maxwell's equations in the presence of an axion background.  We find that the electric field induced by a magnetic field in an axion background is significant and unsuppressed by the size of the experiment.
\end{abstract}

\maketitle
\newpage

\section{Introduction}

Axions and Axion-like particles (ALPs) remain one of the most studied candidates for dark matter. They have gained particular interest in recent years as a discovery of weakly interacting massive particles (WIMPs) has thus far eluded the well-established slate of direct, indirect, and collider probes. An axion that interacts with gluons, that is a QCD axion \cite{Kim:1979if,Shifman:1979if,Dine:1981rt,Zhitnitsky:1980tq}, can solve the strong CP problem while providing a viable dark matter candidate \cite{Preskill:1982cy,Abbott:1982af,Dine:1982ah}.  It has also been shown that ALPs are a natural consequence of certain string theory constructions \cite{Svrcek:2006yi,Arvanitaki:2009fg}, and  ultra-light ALPs have gained particular interest in this context.  For masses around $10^{-20}~\text{eV}$, they could have important consequences for the small-scale structure of the universe and even resolve puzzles, such as the core-cusp \cite{1994Natur.370..629M},  too-big-to-fail \cite{2011MNRAS.415L..40B}, and missing satellites \cite{1993MNRAS.264..201K,1999ApJ...522...82K} problems.

There are several approaches to searching for axions in a laboratory setting~\cite{PhysRevLett.51.1415}, and the best option depends on the particular type of axion-like interactions one introduces. For relatively high masses haloscopes employing microwave cavities, such as the one at the Axion Dark Matter Experiment (ADMX) \cite{ADMX:2018gho} and Center for Axion and Precision Physics Research  CAPP \cite{Adair:2022rtw}, lead the way and cut into the expected QCD axion parameter space.  In haloscope experiments, one searches for microwave excitations produced when a magnetic field in a resonator interacts with the background dark matter axion field. 
 These experiments rely on the electromagnetic interactions of the axion,
\begin{equation}
    \mathcal{L} = g_{a\gamma\gamma} a \mathbf{E} \cdot \mathbf{B},
\end{equation}
where $g_{a\gamma\gamma}$ is the axion-photon coupling and $a$ is the axion field. The theory of an axion with electrodynamics and this interaction is referred to as axion electrodynamics, and it is the foundation of many experimental searches for axions. 

At low masses, for example those of $\mathcal{O}$(10$^{-20}$) eV involved in solving small-scale structure problems, many searches and proposals involve nearly static magnetic fields.  The resulting electric field varies on the period of axion oscillations.\footnote{Nothing about our analysis below changes in the $v \ll 1$ limit if there is a time-dependence to the magnetic field.  There is a modulated periodicity to the resulting electric field with a maximum amplitude suppressed by $m_a / \omega_B$, where $\omega_B$ is the frequency of magnetic field variation.}  There is some controversy surrounding the response electric field in the literature.  The result depends somewhat on the configuration of the system. 
 For cases where a conducing cavity is placed inside a magnetic field, as in the class axion haloscope~\cite{Sikivie:1983ip}, the response electric field is suppressed by $(m_a L)^2$, where $m_a$ is the axion mass and $L$ is the typical size of the cavity. Another interesting case is that of a magnetic field source such as a solenoid or a Helmholtz coil, potentially located in a much larger conducting cavity (such as the Earth's ionosphere/surface or the building containing the laboratory).  Some studies have claimed to find that the response electric field is suppressed by a factor of $(m_a R)^2$, where $m_a$ is the axion mass and $R$ is the typical size of the magnetic field source~\cite{Ouellet:2018nfr,Beutter:2018xfx}.  Other studies have found no such suppression~\cite{Tobar:2018arx} in the classical limit.  The work in Ref.\ \cite{Caldwell:2016dcw} considers a comparable situation where a magnetic field of unspecified origin permeates a dielectric and finds no suppression, though the validity of this should depend on the configuration generating the magnetic field (and any conducting surfaces in the vicinity).  For the frequencies targeted by their proposed experiment, however, this is not a worry as any cavity containing the setup is much bigger than the axion Compton wavelength.  In this work, we aim to definitively clarify this point.\footnote{Situations like that considered in Ref.\ \cite{Arza:2021ekq}, where the magnetic field source is located in a thin ``container'', fall somewhere between the two limits and should be considered separately.}

We derive the response electric field in axion electrodynamics to various magnetic field configurations and in various approximations.  We treat the case of a cavity in a magnetic field more or less as solved, though we briefly review the argument for a suppression by $m_a^2 / \omega_0^2$, where $\omega_0$ is the fundamental frequency of the cavity.  

We focus on a scenario with a relatively isolated magnetic field source with no cavity or least a cavity that is much larger than the source, $L \gg R$.  By carefully considering many aspects of this problem, particularly the solution inside the current carrying wires that are used to generate the zeroth order static magnetic field, we find, in this case, that there is no suppression of the response electric field.  In the region where the magnetic field is applied, our calculations indicate that the electric field goes like
\begin{equation}\label{eq:leading-solution}
    \mathbf{E} \simeq -g_{a\gamma\gamma} [a(t) - a(0)] \mathbf{B}_0,
\end{equation}
where $a(t)$ is the halo axion field, which is approximately uniform over the scale of an experiment, and $\mathbf{B}_0$ is the applied magnetic field.  We assume, without loss of generality, that the otherwise constant magnetic field is turned on at $t = 0$.  Notably, in the limit of perfectly constant axion field, this solution vanishes as one would expect. It means that any experiment looking for this response electric field will have to leave the magnetic field in place for a time of order $1/m_a$ before a signal will be seen.  To leading order in the axion speed $v$, \emph{no} magnetic field response is generated.  This solution is not completely general, but covers a wide range of experimentally-relevant configurations and situations. We further solve the axion electrodynamics equations for different scenarios beyond leading leading approximations and carefully outline and justify the approximations made in all cases. 

There are several ways in which it is interesting to go beyond the leading solution eq.\ \eqref{eq:leading-solution}. First, we determine boundary conditions and  the induced field inside the wires that source the magnetic field in the case of an electromagnet in order to confirm the validity of our solution across space. We then obtain a solution to all orders in the axion velocity $v$, and find both an electric and a magnetic field response. The latter is suppressed by a factor of $v$, but not the size of the field source. We finally consider going beyond leading order in $g_{a\gamma\gamma}$, which, to our knowledge, has not been considered so far. For certain parts of the parameter space, notably regions where $g_{a\gamma\gamma}\mathbf{B}_0/m_{a} \gg$ 1, our calculations indicate that the back-reaction is indeed significant and that care must be taken in determining the validity of a perturbative approach for axion electrodynamics. Importantly, find that for a large laboratory magnetic field of about 10 T, back-reaction becomes relevant for axion masses below $10^{-15}$ eV. 

The remainder of this paper is organized as follows.  In Section \ref{sec:axion-ed}, we review axion electrodynamics and carefully outline the problem we are attempting to solve. Having done so, we solve it in several configurations in Section \ref{sec:solutions}.  We discuss other attempts to solve similar problems in Section \ref{sec:other-work}. Finally, we comment on the physics of our solutions and conclude in Section \ref{sec:conclusions}. We note that we apply the results to this paper in a companion paper Ref.\ \cite{Berger:2022tsn}.

\section{Axion Electrodynamics}\label{sec:axion-ed}

For the purposes of this work, we define axion electrodynamics by the Lagrangian
\begin{equation}
    \mathcal{L} = -\frac{1}{4} F_{\mu\nu}F^{\mu\nu} + \frac{1}{2} \partial_\mu a \partial^\mu a - \frac{1}{2} m_a^2 a^2 - \frac{1}{4} g_{a\gamma\gamma} a F_{\mu\nu} \tilde{F}^{\mu\nu},
\end{equation}
with $\tilde{F}^{\mu\nu} = \epsilon^{\mu\nu\rho\sigma} F_{\rho\sigma}/2$.  From this result, we can derive five coupled equations \cite{PhysRevLett.58.1799}:
\begin{subequations}
\begin{eqnarray}
    \frac{\partial^2 a}{\partial t^{2}} - \nabla^{2}a + m_a^2 a & = & g_{a\gamma\gamma} \mathbf{E}\cdot\mathbf{B}; \\
    \nabla \cdot \mathbf{E} & = & \rho - g_{a\gamma\gamma} \nabla a \cdot \mathbf{B};\\
    \nabla \cdot \mathbf{B} &= & 0; \\
    \nabla \times \mathbf{B} - \frac{\partial \mathbf{E}}{\partial t} & = & \mathbf{J} + g_{a\gamma\gamma} \left(\frac{\partial a}{\partial t} \mathbf{B} + \nabla a \times \mathbf{E}\right);\\\
    \nabla \times \mathbf{E} + \frac{\partial \mathbf{B}}{\partial t} & = & 0, \label{eq:maxwell-axion}
\end{eqnarray}
\end{subequations}
where $\rho$ is the (free) charge density and $\mathbf{J}$ is the steady current sourcing the zeroth order magnetic field.

The solutions to these equations are, in general, non-linear and challenging to solve. As already mentioned in the introduction,  a perturbative approach is possible, but it must be carefully prosecuted---we delay this to Section~\ref{ssec:HigherOrder}. Boundary conditions are another aspect of solving these equations that require particular care.  Far away from the region of our experiment, the applied magnetic field and response electric (and magnetic) field should both vanish.  For the axion, we should then get that far away (but not on, say, galactic scales) the solution reduces an oscillating free axion solution of the form
\begin{equation}
    a = a_0 \cos(E_a t - \mathbf{p}_a \cdot \mathbf{x} + \phi_0),
\end{equation}
where $E_a^2 - |\mathbf{p}_a|^2 = m_a^2$, $\phi_0$ is an arbitrary phase, and the amplitude is of order $a_0 = \sqrt{2 \rho_{\text{DM}}} / m_a$ for a local dark matter density $\rho_{\text{DM}} \approx 0.4~\text{GeV}/\text{cm}^3$.\footnote{As has been emphasized in several recent papers, the amplitude, phase, and velocity change over a coherence time~\cite{Foster:2017hbq,Lisanti:2021vij}.  We assume throughout, as relevant, that the coherence time $\tau_c \sim 1/(m_a \sigma_v^2)$ is much longer than any other time scale in the problem.} It will be convenient to write this axion field far away in a complex presentation,
\begin{equation}
    a^{(0)} = a_0 e^{-i E_a t + i \mathbf{p}_a \cdot \mathbf{x}}.
\end{equation}
This is the zeroth order solution we assume for the axion field in small $g_{a\gamma\gamma}$. Since we are working perturbatively, we are also able to linearize equations~\eqref{eq:maxwell-axion}. For the electric and magnetic fields, we will generally assume that, at zeroth order in $g_{a\gamma\gamma}$, $\mathbf{E}^{(0)} = 0$ and $\mathbf{B}^{(0)} = \mathbf{B}_0$ is some time-independent magnetic field sourced by some thin wires.  Outside of these wires, we assume $\rho = 0$ and $\mathbf{J} = 0$ everywhere.  For the moment, we assume that the entire system is in a vacuum.  Nothing much changes if the system is in a paraelectric material, that is one in which the polarizability is proportional to the unit matrix.  Later we will also consider the case in which the magnetic field is surrounded by a conducting cavity.

We also need to set initial conditions carefully.  Suppose, for example, that we turn on the static magnetic field $\mathbf{B}^{(0)}$ suddenly at $t = 0$.  Then the response fields must be 0 at at $t = 0$.\footnote{We have verified that a continuous, but rapid compared to $1/m_a$ turn on of the magnetic field does not disturb our solution at leading order. This can be verified by modeling the turn on using a signmoid or inverse tan function.}  This can be achieved by adding a electric and magnetic field to cancel out the specific solution we find at $t = 0$, as we show below.  We therefore focus on finding a specific solution for the moment and return to satisfying this initial condition later.

If we find an electric and magnetic field that satisfy Maxwell's equations, along with the initial and boundary conditions outlined above, then the solution is unique.

\section{Solutions to Axion Electrodynamics}\label{sec:solutions}

Having outlined the problem, we now turn to solving it in several limits and approximations.  The simplest one to make is $v \ll 1$.  The halo axion field is expected to be non-relativistic in the local neighborhood, such that
\begin{equation}
    a^{(0)} \simeq a_0 e^{-i m_a t}.
\end{equation}
This limit should be valid when $m_a v R \ll 1$, where $R$ is the typical length scale of the source, provided we can find an $R$-independent solution.  We then study this system beyond the $v = 0$ limit, in fact solving it exactly for the case of an infinite perfect solenoid and providing a vector equation beyond this limit.  Finally, we discuss going beyond leading order in perturbation theory and find that, in some corners of parameter space, the back-reaction on the axion field will be significant and the system is \emph{not} perturbative.

\subsection{The Zero Velocity Limit}

We expand our fields formally in a series
\begin{equation}
    F = F^{(0)} +\epsilon F^{(1)} + \epsilon^2 F^{(2)} + \dots,
\end{equation}
and take $g_{a\gamma\gamma} \to \epsilon g_{a\gamma\gamma}$.  For the moment, we work to $\mathcal{O}(\epsilon)$.  
The equations to this order have the form
\begin{subequations}
\begin{eqnarray}
    \frac{\partial^2 a^{(1)}}{\partial t^{2}} + m_a^2 a^{(1)} & = & 0;\\
    \nabla \cdot \mathbf{E}^{(1)} & = & -g_{a\gamma\gamma} \nabla a^{(0)} \cdot \mathbf{B}^{(0)};\\
    \nabla \cdot \mathbf{B}^{(1)} &= & 0; \\
    \nabla \times \mathbf{B}^{(1)} - \frac{\partial \mathbf{E}^{(1)}}{\partial t} & = & g_{a\gamma\gamma}\left(\frac{\partial a^{(0)}}{\partial t} \mathbf{B}^{(0)}  + \nabla a^{(0)} \times \mathbf{E}^{(0)}\right);\\
    \nabla \times \mathbf{E}^{(1)} + \frac{\partial \mathbf{B}^{(1)}}{\partial t} & = & 0.
\end{eqnarray}
\end{subequations}
Additionally taking the $v \to 0$ limit eliminates all terms involving a gradient of $a^{(0)}$.  For the zeroth order solutions we are assuming, these equations admit a solution
\begin{equation}
    a^{(1)} = 0, \qquad \mathbf{B}^{(1)} = 0, \qquad \mathbf{E}^{(1)} = -g_{a\gamma\gamma} a_0  \mathbf{B}_0 e^{-i m_a t}.
\end{equation}
Note that $\mathbf{B}_0$ is divergence-free (as there are no magnetic monopoles) and curl-free (as there are no sources \emph{locally} for Amp\`ere's law).  This solution is valid except perhaps close to the magnetic field sources, for example some current carrying wires.  

The zeroth order solutions by construction satisfy our boundary conditions.  $\mathbf{E}^{(1)}$ goes to 0 at far away from the magnetic field source since it is proportional to the source's field.  To satisfy the initial conditions, we can add on a space and time-independent field and take the initial conditions to be $\mathbf{B}_0 = 0$ for $t < 0$.  In order to cancel out the solution at $t = 0$ then, we should add a field $g_{a\gamma\gamma} a_0 \mathbf{B}_0$.  With this addition for the purposes of satisfying initial conditions, our solution to first order in $g_{a\gamma\gamma}$ is
\begin{equation}
    a = a_0 e^{-i m_a t}, \qquad \mathbf{B} = \mathbf{B}_0, \qquad \mathbf{E} = - g_{a\gamma\gamma} a_0 \mathbf{B}_0 (e^{-i m_a t} - 1).
\end{equation}

Thus, to the order we are concerned with for the moment, we have constructed a solution that satisfies the boundary conditions and initial condition.  It is therefore \emph{the} solution.

\subsection{Boundary Conditions}

Before moving beyond the $v \to 0$ approximation, we address some potential concerns regarding boundary conditions between regions.  For example, one might wish to consider boundary conditions between a vacuum and a conductor or between two dielectrics.  In order to construct the correct boundary conditions, it is helpful to write~\cite{Tobar:2018arx,Tong:2016kpv}
\begin{equation}
    \mathbf{D} = \mathbf{E} + \mathbf{P} + g_{a\gamma\gamma} a \mathbf{B}, \qquad \mathbf{H} = \mathbf{B} - \mathbf{M} -g_{a\gamma\gamma} a \mathbf{E},
\end{equation}
where $\mathbf{P}$ and $\mathbf{M}$ are the polararization density and magnetization of a material.  As in the case of electrodynamics absent an axion, we can now write Maxwell's equations in the  form
\begin{subequations}
\begin{eqnarray}
		\nabla \cdot \mathbf{D} & = & \rho_f; \\
		\nabla \cdot \mathbf{B}& = & 0; \\
		\nabla \times \mathbf{H}- \frac{\partial \mathbf{D}}{\partial t} & = & \mathbf{J}_f;\\
		\nabla \times \mathbf{E}+ \frac{\partial \mathbf{B}}{\partial t} & = & 0.	
 \end{eqnarray}
 \end{subequations}
and, in this way, recover the standard boundary conditions between media:
 \begin{subequations}
 \begin{eqnarray}
		\hat{n}\cdot (\mathbf{D}_{1} - \mathbf{D}_{2})  & = & \sigma; \label{eq:gauss-bd} \\
		\hat{n}\cdot (\mathbf{B}_{1} - \mathbf{B}_{2})  & = & 0; \label{eq:mag-bd}\\
  		\hat{n} \times (\mathbf{H}_{1} - \mathbf{H}_{2})  & = & \mathbf{K}; \label{eq:ampere-bd} \\
		\hat{n} \times (\mathbf{E}_{1} - \mathbf{E}_{2}) & = & 0,\label{eq:farad-bd}
	\end{eqnarray}
\end{subequations}
where we use 1 and 2 to denote the two media and $\sigma$, $\mathbf{K}$ are the surface charge densities and current densities respectively.

The first boundary to consider is the cavity containing the experiment, whether that be an intentional component of the laboratory setup used to reduce stray field backgrounds or unavoidable conducting boundaries like the building or the Earth's ionosphere/surface.  For a magnetic field source, such as a solenoid or Helmholtz coil, placed inside a larger Faraday cage, we contend that this is not an issue.  The field lines will be distorted close to the conductor walls, but within the solenoid or coils, there should be no concern about this distortion.  This can be easily seen for a box-like conductor, for example, where the usual method of image charges can be applied, at least in the near-static limit and approximating the magnetic field as a dipole field, which then induces a dipole electric field far away from the source.  Image dipoles can be introduced outside the box which, as usual, distort the field lines close to the surface of the box, but leave the field near the center nearly unchanged.

Notably, this differs from the situation where a conducting cavity is placed inside a uniform magnetic field.  In that case, compatibility with the boundary conditions, that the electric field parallel to the conductor walls vanishes, enforces that the electric field in the \emph{entire} cavity is suppressed by $m_a^2 / \omega_0^2$ where $\omega_0$ is the fundamental mode of the cavity and $m_a \ll \omega_0$.  As $m_a \to \omega_0$, resonances are excited and we reach the regime of traditional haloscope experiments~\cite{PhysRevLett.51.1415}.  Other recent haloscope proposals such as \cite{Lawson:2019brd} also involve conducting cavities in a uniform magnetic field and thus would also see a suppression in the small frequency limit we mostly consider in this work.

To see this, we apply the usual expansion in modes for the vector potential inside a resonator
\begin{equation}
    \mathbf{A} = \sum_\lambda q_\lambda(t) \mathbf{A}_{\lambda}(\mathbf{x}),
\end{equation}
where $\lambda$ label the modes of the cavity satisfying the correct boundary conditions.  $\mathbf{A}_\lambda$ are the appropriately normalized mode functions.  The equation for $q_\lambda$ is
\begin{equation}
    \ddot{q}_\lambda + \omega_\lambda^2 q = \int dV \, \mathbf{J}_{\text{eff}} \cdot \mathbf{A}_\lambda.
\end{equation}
The effective current density here can be read off of eq.\ \eqref{eq:maxwell-axion}, that is
\begin{equation}
    \mathbf{J}_{\text{eff}} = -i m_a g_{a\gamma\gamma} a_0 \mathbf{B}_0 e^{-i m_a t}.
\end{equation}
When resonances are a possibility, we can include a friction term to model this~\cite{Sikivie:2020zpn}, but we are most concerned with the region far from resonance in the ultralight regime.  The specific solution to this equation is
\begin{equation}
    q_\lambda = \frac{1}{\omega_\lambda^2 - m_a^2} \int dV \, \mathbf{J}_{\text{eff}} \cdot \mathbf{A}_\lambda
\end{equation}
The remainder of the calculation depends on the specific shape of the cavity.  In any case, provided that the magnetic field is homogeneous in the cavity, the fundamental mode will typically dominate.  We can estimate the solution as
\begin{equation}
    |\mathbf{A}| \sim \frac{m_a g_{a\gamma\gamma} a_0  V^{3/2} |\mathbf{B}_0  \cdot \mathbf{A}_0|}{\omega_0^2} e^{-i m_a t},
\end{equation}
where $V$ is the volume of the cavity.  The $V^{3/2}$ arises due to the required normalization of the mode functions $\mathbf{A}_\lambda$.  In other words, $V^{3/2} \mathbf{A}_0$ will be dimensionless and order 1 in the middle of the volume.  From this solution, we can estimate the electric field by taking a time derivative,
\begin{equation}
    |\mathbf{E}| \sim \frac{m_a^2}{\omega_0^2} g_{a\gamma\gamma} a_0 |\mathbf{B}_0|.
\end{equation}

One way to see the difference in this case is that the response field generated by the surface charges on the conductor must arrange to cancel the parallel field there. For a constant magnetic field, since the axion-induced field at the boundary is the same as the field in the center, this mostly cancels out the field at the center as well.  On the other hand, if the magnetic field source is located in the middle of the cavity, then the boundary magnetic field is much smaller than the field in the center.  The field induced by the charges is thus small compared to the field inside the source.


Similarly, one might worry about the boundary conditions at the wires of an electromagnet or permanent magnet sourcing the magnetic field. This situation is far more complicated to deal with.  We demonstrate here that the field outside  long, straight current carrying wires is \emph{not} in fact distorted at all.  We assume that the wires carry a constant current density $\mathbf{J}_0$ generated according to Ohm's law by an electric field $\mathbf{E}_0 = \mathbf{J}_0 / \sigma$, where $\sigma$ is the conductivity.  Furthermore, the magnetic field profile inside the wire is sourced by the current density such that Ohm's law holds:
\begin{equation}
    \nabla \times \mathbf{B}_0 = \mu \mathbf{J}_0 = \frac{\mu}{\sigma} \mathbf{E}_0.
\end{equation}
Note that for a constant current, by the continuity equation, $\nabla \cdot \mathbf{J}_0 = 0$ and, by Ohm's law and Faraday's law, $\nabla \times \mathbf{J}_0 = 0$.  Then, the first order Maxwell's equations inside the conducting wires can be written as
\begin{subequations}
    \begin{eqnarray}
    \nabla \cdot \mathbf{E}^{(1)} & = & -g_{a\gamma\gamma} \nabla a^{(0)} \cdot \mathbf{B}^{(0)};\\
    \nabla \cdot \mathbf{B}^{(1)} &= & 0; \\
    \nabla \times \mathbf{B}^{(1)} - \mu \epsilon \frac{\partial \mathbf{E}^{(1)}}{\partial t} & = & \sigma \mathbf{E}^{(1)} + g_{a\gamma\gamma}\left(\frac{\partial a^{(0)}}{\partial t} \mathbf{B}^{(0)}  + \nabla a^{(0)} \times \mathbf{E}^{(0)}\right);\\
    \nabla \times \mathbf{E}^{(1)} + \frac{\partial \mathbf{B}^{(1)}}{\partial t} & = & 0.
\end{eqnarray}
\end{subequations}
Under the assumed conditions of a constant current (as well as constant material properties of the wire), we can write a solution to these equations (to zeroth order in axion velocity) as
\begin{subequations}
\begin{eqnarray}
\mathbf{E}^{(1)} & = & \frac{i \omega_a}{\sigma -i \epsilon \omega_a} g_{a\gamma\gamma} a^{(0)} \mathbf{B}_0, \\
\mathbf{B}^{(1)} & = & \frac{\mu}{\sigma -i \epsilon \omega_a} g_{a\gamma\gamma} a^{(0)} \mathbf{J}_0.
\end{eqnarray}
\end{subequations}
We need to check the consistency of these solutions with the boundary conditions at the interface between the wire and the vacuum outside.  

Both the electric and magnetic fields generated at first order are parallel to the wire surface.  For the magnetic field, any mismatch between the solution inside and outside the wire can be accounted for by a surface current.  

For the parallel component of the electric field, in deriving a condition based on Faraday's law, we must be careful.  There is potentially a divergence in the magnetic field across the boundary, since $\mathbf{J}_0 = \nabla \times \mathbf{B}_0 / \mu$.  There is a jump in the magnetic field across the boundary if there is a non-trivial permeability $\mu$, which leads to a divergence in the curl of the magnetic field.  Taking this into account, we find that taking a small loop across the boundary, the full boundary condition,
\begin{equation}
\oint \mathbf{E} \cdot d\boldsymbol{\ell} + \int \frac{\partial \mathbf{B}}{\partial t} \cdot d\mathbf{A} = 0,
\end{equation}
is satisfied after applying Stokes' theorem on the second term, which goes like $\mathbf{J}_0$.  Thus, we find that although the electric field inside the wire may be very small, suppressed by $\omega_a/\sigma \ll 1$, it need not be small outside the wire.\footnote{We note that Tobar et al. reach a similar conclusion in \cite{Tobar:2018arx} using an impressed current formalism typically favoured in engineering applications of electrodynamics.}

Finally, we comment on the common example of an infinite solenoid that is considered in the literature~\cite{Tobar:2018arx,Ouellet:2018nfr,Beutter:2018xfx}. This system can seem confounding at first, as one might think to apply the boundary conditions \eqref{eq:farad-bd} directly from the inside solution to an outside solution.  This procedure is not correct, as one needs to go through the wires.  One can attempt to apply the boundary conditions more carefully, but there is a simpler way to get the correct field given our general solution.  We solve for the fields induced by a finite solenoid take the infinite length limit.  The response field outside the solenoid goes to 0 as the length of the solenoid becomes large as the magnetic field outside the solenoid vanishes.  Thus, for an infinite solenoid that generates a magnetic field $\mathbf{B}_0$ within the coils, one finds
\begin{equation}
    \mathbf{E}^{(1)} =\left\{\begin{array}{l l} - g_{a\gamma\gamma} a_0 \mathbf{B}_0 (e^{-i m_a t} - 1), & r < R; \\
    0, & r > R.
    \end{array}\right.,
\end{equation}
where $r$ is the distance away from the solenoid axis and $R$ is the solenoid radius. 

\subsection{All orders in velocity}
In this section, we consider corrections to our solution due to finite speed and demonstrate that these are, as expected, small in general. As in the previous section, we assume the source magnetic field $\mathbf{B}_0$ is independent of time.  To obtain a solution to all orders in $v$, we can start by noting that we can write Amp\`ere's law as
\begin{equation}\label{eq:formal-ampere}
    \nabla\times \mathbf{B}^{(1)} + i E_a \mathbf{E}^{(1)} = -i E_a g_{a\gamma\gamma} a^{(0)} \mathbf{B}_0,
    \end{equation}
assuming that $\mathbf{E}^{(1)}$ has the required $e^{-i E_a t}$ time dependence.  In other words, we have the formal solution:
\begin{equation}
    \mathbf{E}^{(1)} = -g_{a\gamma\gamma} a^{(0)} \mathbf{B}_0 -\frac{1}{i E_a} \nabla \times \mathbf{B}^{(1)}.
\end{equation}
This ansatz automatically solves Gauss' law with the added axion term as the divergence of the second term vanishes and the divergence of the first term gives the right-hand side of Gauss' law. 

We can take a curl of this solution to find
	\begin{equation}
		\nabla \times \mathbf{E}^{(1)} = -g_{a\gamma\gamma} \nabla a^{(0)} \times \mathbf{B}_0 +\frac{1}{i E_a} \nabla^2 \mathbf{B}^{(1)},
  \end{equation}
  using Gauss' law for magnetism to simplify the second curl.  Plugging this into Faraday's law to get a differential equation for $\mathbf{B}^{(1)}$:
  \begin{equation}
    \nabla^2 \mathbf{B}^{(1)} + E_a^2 \mathbf{B}^{(1)} = E_a g_{a\gamma\gamma} a^{(0)} \mathbf{p}_a \times \mathbf{B}_0.
  \end{equation}
In principle, this equation can be simply solved using the well-known Green's function(s) for the LHS operator, but the required integration here may be non-trivial and depends on the choice of magnetic field profile.

Rather than construct this solution for some rather complicated magnetic field profiles, we focus for now on the solenoid, using the same geometry as in the previous section.  If the source magnetic field has no spatial dependence, as with a solenoid, then we know that
\begin{equation}
    \mathbf{B}^{(1)} = e^{-i E_a t +i \mathbf{p}_a \cdot\mathbf{x}} \mathbf{b},
\end{equation}
for some constant vector $\mathbf{b}$ in order to match the spatial dependence arising exclusively from $a^{(0)}$.  We can simply plug in to Faraday's law and solve for the magnetic field:
\begin{equation}
    \mathbf{B}^{(1)} =  \gamma_a^2 g_{a\gamma\gamma} a^{(0)} \mathbf{v}_a \times \mathbf{B}_0,
\end{equation}
where $\gamma_a = E_a / m_a$ and $\mathbf{v}_a = \mathbf{p}_a / E_a$ are the boost factor and velocity of the axion field.  Plugging back into our formal Amp\'ere's law solution \eqref{eq:formal-ampere} gives
\begin{equation}
    \mathbf{E}^{(1)} = -g_{a\gamma\gamma} a^{(0)} \mathbf{B}_0 + \mathbf{v}_a \times \mathbf{B}^{(1)}.
\end{equation}
Both solutions reduce to the $v \to 0$ solutions in that limit and, as with the zero-velocity approximation, the induced fields outside the solenoid vanish in this limit.

One challenge in this particular case is satisfying the initial conditions.  In the zero-velocity limit, we could simply add a constant field, which automatically satisfies the vacuum solutions, such that the field vanishes at $t = 0$.  In this section, we cannot add a constant to the solution as it has spatial dependence at t = 0.  

What's more, in this case, the correct initial condition is not that the first-order fields vanish at $t = 0$.  Rather, we note that turning on the current that sources the magnetic field generates a time-dependent magnetic field, which in turn generates a transient electric field.  This transient electric field enters our equations.  Integrating over the small time window during which the magnetic field is turned on leaves a small imprint on the induced magnetic field even after the steady state is reached.  By modeling the turn on of the current by a Heaviside $\theta$ function, we find a $\delta$ function induced electric field at 0th order.  Integrating over a small time window around $t = 0$, we find an initial condition
\begin{equation}
    \mathbf{E}^{(1)}(t = 0^+) =  - g_{a\gamma\gamma} \nabla a(t = 0) \times \left[-\frac{|\mathbf{B}_0|}{2} (-y \hat{x} + x \hat{y})\right].
\end{equation}
We see that this effect vanishes as $v \to 0$.  We do not attempt to find a solution consistent with this initial condition.

The conclusion of this section is that there is no barrier to solving axion electrodynamics in the presence of a full plane wave axion field.  We obtain solutions that reduce to the simpler $v \to 0$ solutions in that limit. Before turning to a discussion of these results, we comment on higher order corrections in $g_{a\gamma\gamma}$.


 \subsection{Higher Order Corrections in $g_{a\gamma\gamma}$}\label{ssec:HigherOrder}
The last major approximation that we've made in this calculation is to work to linear order in $g_{a\gamma\gamma}$.  This approximation is crucial as higher order corrections in this variable imply potential sensitivity to higher order terms in the axion effective field theory.  On the other hand, it is not entirely obvious that this approximation is under control.  In this section, we show that there are indeed relevant parts of parameter space at sufficiently large magnetic field where the perturbative approximation we have made breaks down.

We've already seen in the previous sections the solution for a solenoid to 0th order and 1st order in $g_{a\gamma\gamma}$.  Here, we work in the $v \to 0$ limit to second order in $g_{a\gamma\gamma}$.  We solve only inside the solenoid; there are complications to solving outside as we will see.  In any case, what we would like to understand is when the higher order terms are relevant at all.  If they are relevant, we should not pursue our perturbative calculation any further. 

At the second order in perturbation theory, inside the solenoid, as we have seen, there is a non-trivial $\mathbf{E}^{(1)}$ and $\mathbf{B}^{(1)}$ that contributes, in addition to $\mathbf{B}^{(0)}$ and $a^{(0)}$.  Keeping all terms quadratic in $g_{a\gamma\gamma}$, we get the following equations
\begin{subequations}
 \begin{eqnarray}
     \Box a + m_a^2 a & = &  g_{a\gamma\gamma}^2 \mathbf{E}^{(1)} \cdot \mathbf{B}^{(0)} \label{eq:axion-2nd-order} \\
    \nabla \cdot \mathbf{E}^{(2)} & = & -g_{a\gamma\gamma} \nabla a^{(0)} \cdot \mathbf{B}^{(1)} \\
    \nabla \cdot \mathbf{B}^{(2)} & = & 0 \\
    \nabla \times \mathbf{B}^{(2)} - \frac{\partial \mathbf{E}^{(2)}}{\partial t} & = &  g_{a\gamma\gamma} \frac{\partial a^{(0)}}{\partial t} \mathbf{B}^{(1)} + g_{a\gamma\gamma} \nabla a^{(0)} \times \mathbf{E}^{(1)} \\
    \nabla \times \mathbf{E}^{(2)} + \frac{\partial \mathbf{B}^{(2)}}{\partial t} & = & 0. 
\end{eqnarray}
\end{subequations}
Note first that in the limit of $v \to 0$, the electric and magnetic fields are unaffected by this perturbation: both $\mathbf{B}^{(1)}$ and $\nabla a^{(0)}$ vanish for $v \to 0$.  

Our goal then is to solve \eqref{eq:axion-2nd-order}.  For notational clarity, we will drop the superscript $(2)$ at this point.  We will also switch to a real representation of the axion field.  Neglecting any spatial dependence in the uniform magnetic field and $v \to 0$ limit, the equation we are solving is
\begin{equation}
    \frac{\partial^2 a^{(2)}}{\partial t^2} + m_a^2 a^{(2)} = - g_{a\gamma\gamma}^2 B_0^2 a^{(0)} =  - g_{a\gamma\gamma}^2 B_0^2 a_0 \cos(m_a t + \phi_0),
\end{equation}
where $\phi_0$ is an arbitrary phase.  This is the equation of a harmonic oscillator driven at its resonant frequency, which leads to secular growth.  It is not strictly necessary to eliminate the secular growth term in order to determine the scaling of this solution, but we do so anyway.

To do this, we work with the effective Hamiltonian
\begin{equation}
   H = \frac{1}{2} \pi^2 + \frac{1}{2} m_a^2 a^2 +  g_{a\gamma\gamma} a \mathbf{E}\cdot \mathbf{B},
\end{equation}
with $\pi$ the canonical momentum corresponding to $a$.
We define the 0th order Hamiltonian
\begin{equation}
    H_0 = \frac{1}{2} \pi^2 + \frac{1}{2} m_a^2 a^2.
\end{equation}
Next, we can do a canonical transformation to action to action angle variables,
\begin{equation}
    I = \frac{1}{2 \pi} \oint \pi da.
\end{equation}
At 0th order, we can solve for $I$ for any trajectory 
\begin{equation}
    a = a_0 \cos(m_a t + \phi_0).
\end{equation}
Using conservation of energy, we can write
\begin{equation}
    \pi = \sqrt{2 E - m_a^2 a^2},
\end{equation}
which we can integrate over a 0th order period of motion to find, for a motion with total energy $m_a^2 a_0^2 / 2$,
\begin{equation}
    I = \frac{m_a a_0^2}{2}.
\end{equation}
Plugging all of this back into the Hamiltonian, we have
\begin{equation}
    H_0 = m_a I,
\end{equation}
which gives the equations of motion
\begin{equation}
    \dot{I} = 0, \qquad \dot{\phi} = m_a,
\end{equation}
with $\phi$ being the angle variable canonically conjugate to $I$.  We can thus identify
\begin{equation}
    \phi = m_a t + \phi_0,
\end{equation}
and so
\begin{equation}
    a = \sqrt{\frac{2 I}{m_a}} \cos\phi.
\end{equation}
Next, plugging this back into the full Hamiltonian, we have
\begin{equation}
    H = m_a I + g_{a\gamma\gamma} \sqrt{\frac{2 I}{m_a}}  \cos\phi \mathbf{E}\cdot\mathbf{B}.
\end{equation}
The equations of motion here are
\begin{equation}
    \dot{\phi} = \frac{\partial H}{\partial I} = m_a + g_{a\gamma\gamma} \frac{1}{\sqrt{2 m_a I}} \cos\phi\mathbf{E} \cdot \mathbf{B},\qquad \dot{I} = - \frac{\partial H}{\partial \phi} = g_{a\gamma\gamma}  \sqrt{\frac{2 I}{m_a}}\sin\phi \mathbf{E}\cdot\mathbf{B}.
\end{equation}

The 0th order equations are solved by
\begin{equation}
    \phi^{(0)} = m_a t + \phi_0,\qquad I^{(0)} = \frac{m_a a_0^2}{2}.
\end{equation}
We can write the next term in the equation as
\begin{equation}
    \dot{\phi}^{(1)} = g_{a\gamma\gamma} \frac{1}{\sqrt{2 m_a I^{(0)}}} \cos\phi^{(0)}\mathbf{E} \cdot \mathbf{B}, \qquad \dot{I}^{(1)} = g_{a\gamma\gamma}  \sqrt{\frac{2 I^{(0)}}{m_a}}\sin\phi^{(0)} \mathbf{E}\cdot\mathbf{B}.
\end{equation}
To leading order, we have 
\begin{equation}
    \mathbf{E}\cdot\mathbf{B} = -g_{a\gamma\gamma} B_0^2 a^{(0)} =  \boldsymbol{-}g_{a\gamma\gamma} B_0^2  \sqrt{\frac{2 I^{(0)}}{m_a}} \cos\phi^{(0)}.
\end{equation}
Plugging this in, along with our solutions for $I^{(0)}$ and $\phi^{(0)}$, we have equations 
\begin{equation}
     \dot{\phi}^{(1)} = - \frac{g_{a\gamma\gamma}^2 B_0^2}{m_a} \cos^2(m_a t + \phi_0),\qquad  \dot{I}^{(1)} = - g_{a\gamma\gamma}^{2} a_0^2 B_0^2 \cos(m_a t + \phi_0) \sin(m_a t + \phi_0). 
\end{equation}
These equations can be solved, giving (if we take $\phi_0 = 0$ WLOG): 

\begin{equation}
    \phi^{(1)} = -\frac{g_{a\gamma\gamma}^2 B_0^2}{4 m_a^2} [2 m_a t + \sin(2 m_a t)],\qquad I^{(1)} =   \frac{g_{a\gamma\gamma}^2 B_0^2 a_0^2}{4 m_a^2} \cos(2 m_a t), 
\end{equation}
which we can use to get our perturbed $a$ solution: 
\begin{equation}
    a = \sqrt{\frac{2 I}{m_a}} \cos \phi =  a_0\left(1 + \frac{g_{a\gamma\gamma}^2 B_0^2}{2 m_a^{2}}  \cos^2(m_a t)\right) \cos\left(m_a t + \frac{g_{a\gamma\gamma}^2 B_0^2}{4 m_a^2} [2 m_a t + \sin(2 m_a t)] \right).
\end{equation}
A couple of notes are in order here.  First, one can check explicitly that this result solves the equations of motion up to order $g_{a\gamma\gamma}^2$.  Second, the corrections to both the amplitude and frequency may non-perturbatively large.  These are all of order
\begin{equation}
    \frac{g_{a\gamma\gamma}^2 B_0^2}{m_a^2} \approx 4 \times \left(\frac{g_{a\gamma\gamma}}{10^{-10}~\text{GeV}^{-1}}\right)^2 \left(\frac{B_0}{1~\text{mT}}\right)^2 \left(\frac{m_a}{10^{-20}~\text{eV}}\right)^{-2}.
\end{equation}
The problem of non-perturbativity is relevant for masses close to the bounds set by small scale structure (for example using the Lyman-$\alpha$ forest \cite{Rogers:2020ltq}) at relatively large magnetic fields. By a mass of about $10^{-15}~\text{eV}$ or higher, concerns about perturbativity are not relevant for any magnetic field up to $10~\text{T}$, around the limit of what might reasonably be used with current technology.

\section{Comparison with Other Work}
\label{sec:other-work}

Several other works have attempted to solve the equations of axion electrodynamics, especially in the context of a zeroth order magnetic field sourced by an infinite solenoid.  Some have agreed with our result in the $v \to 0$ limit~\cite{Tobar:2018arx}, but others have found an $(m_a R)^2$ suppression of the induced electric field (R here is the solenoid radius), along with a magnetic field that goes like $m_a R$.  Two works in particular~\cite{Ouellet:2018nfr,Beutter:2018xfx} have attempted using different techniques to construct an explicit solution for certain geometries and approximations, under conditions similar to those assumed in this work.  We address some concerns with these works in this section.

We begin with the work by Ouellet and Bogorad~\cite{Ouellet:2018nfr}.  The authors take derivatives of the equations for axion electrodynamics to write a wave equation. Notably, Maxwell's equations are not the wave equation; they imply the wave equation, but the two are not equivalent. We focus on their solution for an infinite solenoid. The authors find a non-zero (but suppressed) axial $E$ field with radial profile $\psi_E$ and a non-zero angular $\mathbf{B}$ with radial profile $\psi_B$ and further claim matching conditions at the boundary on $\psi_E$, $\psi_B$, and their derivatives, as one would do for a second order differential equation.  We find this is not the case for multiple reasons.  First, we are able to explicitly construct a solution for a finite solenoid, from which we find that the field outside the solenoid should follow the magnetic field. In other words, since the magnetic field vanishes outside the solenoid, so does the electric field. The reason is that we can't directly apply the matching conditions to the fields inside and outside the solenoid---we must consider the effects of the wires of the solenoid here.

Further, adopting their notation, one should not in general apply the conditions to $\psi_E^\prime$ and $\psi_B^\prime$.  Rather, one should determine from Maxwell's equations a consistent relation for the derivatives that must be satisfied everywhere:
\begin{subequations}
\begin{eqnarray}
    \frac{1}{\rho} \frac{d}{d \rho} (\rho \psi_B) - i \omega_a \psi_E & =  & J_\phi, \\
    i \omega_a \psi_B - \frac{d \psi_E}{d\rho} = 0.
\end{eqnarray}
\end{subequations}
where $J_\phi$ is the effective current density in the $\phi$ direction.  This may be discontinuous across a boundary, but as we go through in other parts of this work, there are subtleties crossing a current-carrying wire anyway.

The authors also consider a polarization approach.  The solution for the polarization that they write in equation (48) is indeed correct. But even though there is no divergence, there is no reason that should make it unphysical.  It has time dependence and so is not just a constant term. It is clearly consistent with (40d) if we set $\mathbf{D} = 0$.  


We also consider the work of Beutter et.\ al.\ \cite{Beutter:2018xfx}. The authors employ a quantum field theory approach.  First, we note that the approach that they take is actually entirely classical: it is nothing but solving the vector potential version of Maxwell's equations using Green's function methods. This is a completely valid approach to solving this problem, but one needs to be careful with what Green's function is used.  Following usual quantum field theory logic, the authors use the Feynman Green's function.  On the other hand, one should be careful. The correct Green's function to use here is the retarded Green's function.  The Feynman Green's function is acausal for the classical field configuration background that is being applied in this case.

A further sign of issues with the fields they generate in the solenoid case is that their $\mathbf{E}$-field solution does not satisfy the axion electrodyanmics version of Maxwell's equations, particularly Amp\`ere's law.  A Green's function approach to this problem renders it far more complicated than is necessary.  At the end of the day, the interior solution is already an eigenfunction of the d'Alembertian operator.  Outside the solenoid, the d'Alembertian vanishes and we are encountering a 0 eigenvalue.  In such a case, it is not necessary to go through the trouble of inverting the operator and dealing with these subtleties.  We do not pursue this approach further.

Ultimately, the test of a solution to axion electrodynamics should be whether it satisfies the axion electrodynamics equations of motion, along with appropriate initial conditions and boundary conditions.  In this work, we have constructed solutions that do so, up to the perturbative approximation.  We validated the consistency of perturbation theory and carefully outlined where it breaks down. 

\section{Discussion and Conclusions}
\label{sec:conclusions}

In this work, we have shown that in the presence of a static magnetic field source in an arbitrary geometry and a slowly evolving plane-wave axion field, there is a solution for the electric field of the form
\begin{equation}
    \mathbf{E} = -g_{a\gamma\gamma} [a(t) - a(0)] \mathbf{B}.
\end{equation}
This solution is valid everywhere except near a conducting surface or inside a current-carrying wire.  It has been shown to satisfy boundary and initial conditions.  It vanishes in the case of $m_a \to 0$.  It can be extended to the interior of current-carrying wires and to all orders in speed.

It has been widely claimed~\cite{Ouellet:2018nfr,Beutter:2018xfx,Arza:2021ekq} that the axion solution should be suppressed by $(m_a R)^2$.  We find that this is not the case for a magnetic field source inside a large ``cavity.''  

This solution is somewhat unintuitive from the point of view of considering the axion term in Amp\`ere's law as an effective current density.  We typically think of the curl of the magnetic field as being the dominant term in Amp\'ere's law outside the ultra-relativistic limit, but we are presented here with a situation that is quite different from what is generally set up using standard model physics.  In the work here, we are examining fields generated \emph{inside} an effective current density. We end up with an electric field rather than a magnetic field at leading order.

This work motivates a renewed effort to search for electric field responses in the presence of a magnetic field.  The ``smoking gun'' for the electric field being ultralight axion-induced is a slow variation of the amplitude of that electric field.  The electric field response starts at zero, so the magnetic field must be left on a time of order an axion period before any effect is seen.  In a companion paper, we demonstrate that current spectroscopic measurements are sensitive to such an electric field response, but one could imagine optimizing these experiments further to gain even better sensitivity to axion dark matter.  By probing the system at a series of increasing magnetic fields, one could easily evade constraints on the validity of the perturbative approach in this work.

{\bf Acknowledgments.} We thank Javier Acevedo, Brian Batell, Michael Fedderke, Julia Gehrlein, Marty Gelfand, Saarik Kalia, and Pranava Teja Surukuchi, for useful discussions. This material is based upon work supported by the National Science Foundation under Grant No.\ 2112789.

\bibliographystyle{unsrt}
\bibliography{apssamp}

\end{document}